\documentclass[nofootinbib,superscriptaddress]{revtex4}
\pdfoutput=1
\usepackage{amsmath,amssymb}
\usepackage{epsfig}
\usepackage{hyperref}
\usepackage{color}
\usepackage{slashed}
\usepackage{placeins}

\usepackage{amsfonts}
\usepackage{graphicx, rotating}
\usepackage{epstopdf}
\usepackage{latexsym}
\usepackage{rotating}

\usepackage[export]{adjustbox}

\usepackage[font={small}]{caption}   

\definecolor{myred}{rgb}{0.7, 0, 0}
\definecolor{myblue}{rgb}{0, 0, 0.7}
\definecolor{mygreen}{rgb}{0.04, 0.7, 0.5}

\hypersetup{colorlinks,citecolor=myred,linkcolor=myblue,urlcolor=myblue,linktocpage=true}

 \def\be   {\begin{equation}}   \def\ee   {\end{equation}}
 \def\ba   {\begin{array}}      \def\ea   {\end{array}}
 \def\bea  {\begin{eqnarray}}   \def\eea  {\end{eqnarray}}
 \def\bean {\begin{eqnarray*}}  \def\eean {\end{eqnarray*}}
 \def\nn{\nonumber}
 \def\bry{\begin{array}}
 \def\ery{\end{array}}

\newcommand{\hhref}[1]{\href{http://arxiv.org/abs/#1}{arXiv:#1}}

\numberwithin{equation}{section}

\begin{document}

\title{
Mirror Cosmological Relaxation of the Electroweak Scale
}

\author{Oleksii Matsedonskyi}
\email{oleksii.matsedonskyi@sns.it}
\address{Scuola Normale Superiore, Piazza dei Cavalieri 7, 56126 Pisa, Italy}

\begin{abstract}
The cosmological relaxation mechanism proposed in~\cite{Graham:2015cka} allows for a dynamically generated large separation between the weak scale and a theory cutoff, using a sharp change of theory behaviour upon crossing the limit between unbroken and broken symmetry phases. 
In this note we present a variation of this scenario, in which stabilization of the electroweak scale in the right place is ensured by the $Z_2$ symmetry exchanging the Standard Model (SM) with its mirror copy. 
We sketch the possible ways to produce viable thermal evolution of the Universe and discuss experimental accessibility of the new physics effects.  
We show that in this scenario the mirror SM can either have sizeable couplings with the ordinary one, or, conversely, can interact with it with a negligible strength. 
The overall cutoff allowed by such a construction can reach $10^9$~GeV.
\end{abstract}

\maketitle





\section{Introduction}

The Gauge Hierarchy problem has motivated development of the new physics models allowing to screen the sensitivity of the SM to the physical mass scales much above the electroweak (EW) scale. In most of known examples of such scenarios (e.g. supersymmetry, compositeness, extra dimensions) the new physics responsible for such a screening unavoidably has EW charges and appears not far above the EW scale. This feature is absent in the new approach to the problem presented in~\cite{Graham:2015cka}
\footnote{See~\cite{Espinosa:2015eda,Hardy:2015laa,Patil:2015oxa,Antipin:2015jia,Jaeckel:2015txa,Gupta:2015uea,Batell:2015fma,Kobakhidze:2015jya} for subsequent developments and related works. See also~\cite{Dvali:2003br,Dvali:2004tma} for previous works suggesting Higgs mass scanning during cosmological evolution.} , where the small value of the electroweak scale is chosen dynamically, using the fact that the theory undergoes qualitative changes upon crossing the $m_h^2 = 0$ point. Technically, in this case the Higgs mass receives a contribution from an additional scalar  field $\phi$ (``relaxion"), such that 
\be \label{eq:mh}
m_h^{2} = M^2 - g \phi \, ,
\ee
where $M$ is a cutoff-size Higgs mass and $g$ is a dimension-one coupling which breaks the axion shift symmetry and hence can be naturally small. The relaxion rolls down its potential 
\be
V_\phi = - g M^2 \phi \,,
\ee 
whose slope is also controlled by $g$, and hence
performs a scan of $m_h^2$ values, by assumption starting at some $m_h^2>0$. The excess of the axion energy released while moving down the potential is dispersed due to a Hubble friction (see~\cite{Hardy:2015laa} for an alternative way of scanning). 
Once $m_h^2$ crosses the zero point and the Higgs field develops a vacuum expectation value (VEV), another part of the relaxion potential should generate the potential barriers and stop the relaxion. This second part of the potential can be generically written as~\cite{Espinosa:2015eda}
\be \label{eq:bar}
V_{\text{B}}(h) = \epsilon ( c_0 \Lambda^4 + c_1 \Lambda^3 h +c_2 \Lambda^2 h^2 + ...) \cos \phi /f \,,
\ee
where $\Lambda$ is some characteristic mass scale of the physics which generates the barriers and $f$ is a relaxion scale. The potential $V_{\text{B}}$ contains a weaker breaking of the relaxion shift symmetry and hence its size, controlled by $\epsilon$, is independent of $g$. Not all the terms listed in~(\ref{eq:bar}) are necessarily always generated, e.g. the terms linear in $h$ require additional EW symmetry breaking (EWSB) sources to participate in the formation of the barriers.  
Adjusting the slope of the $V_\phi+V_{\text{B}}$ potential one can arrange it to become zero when $V_{\text{B}}(h) = V_{\text{B}}(v_{\text{SM}})$, hence trapping $\phi$ in a minimum with the desired Higgs VEV. One of the necessary conditions for this mechanism to be natural is a smallness of the first term in the potential~(\ref{eq:bar}), which is independent of the Higgs field and hence does not change upon EWSB. With this term being dominant,  the peculiarity of the $m_h^2=0$ point would be washed away. Nevertheless EWSB-independent terms are expected to be generated radiatively from the Higgs-dependent ones.
One of the ways to preserve dominance of the Higgs-dependent terms is to have $\Lambda^4$ around or below the EW scale, this is the case of the models presented in~\cite{Graham:2015cka}. This smallness of $\Lambda$ in general does not require the overall cutoff $M$ to be small as well. An alternative way was proposed in~\cite{Espinosa:2015eda}, where the constant coefficient $c_0$ is promoted to a field-dependent one, in the same way as the Higgs mass in Eq.~(\ref{eq:mh}). This allows to decrease $c_0$ to an acceptable size during the evolution of the additional scanning field.

In this work we discuss one more way to generate viable  potential barriers. 
In order to protect the barriers from quantum corrections we will use a weakly broken $Z_2$ symmetry between the Standard Model and its hypothetical mirror copy~\footnote{See~\cite{Okun:2006eb} for a review of the mirror models development.}. This will ensure the cancellation of large corrections to the oscillatory term, which is assumed to have a $Z_2$-breaking form
\be\label{eq:bar2}
m^2 (h_1^2-h_2^2) \, \cos \phi / f \,.
\ee
The mirror copy of the Standard Model is supposed to interact with the SM only by means of the scanning field $\phi$, i.e. the two sectors 
do not share any internal gauge symmetries.
As we will show in the following, despite the presence of additional light states at and below the electroweak scale, they can be negligibly weakly coupled to the SM, not leaving chances to be observed in the current or far future collider experiments, while the overall cutoff of the theory can reach $10^9$~GeV.   

Throughout all the discussion we will assume the inflation going on in background, providing a Hubble friction needed for the scanning, but we will not give any concrete details of the inflation sector itself, limiting ourselves to pointing out the constraints imposed by the relaxation on the inflation and a subsequent reheating.

In Section~\ref{sec:main} we describe the mirror relaxation mechanism, 
the constraints imposed on the inflation allowing the relaxation to work are summarized in Section~\ref{sec:consist}; in Section~\ref{sec:thermhist} we discuss the ways to reconcile the presence of the mirror SM with the cosmological observations, and estimate the resulting bounds on the strength of interactions between the two SM's; Section~\ref{sec:disc} contains a discussion of our results and conclusions.      

\section{Central Mechanism}
\label{sec:main}

In the first approximation our model is described by two non-interacting analogues of the Standard Model, called in the following SM$_1$ and SM$_2$. The leading source of communication between the two sectors is the scalar field $\phi$, singlet under both SM$_1$ and SM$_2$ gauge symmetries. The Higgs fields of both sectors are coupled in a symmetric way to the field $\phi$, whose VEV enters the effective Higgs masses as $m_h^2 = M^2-g \phi$, where $g$ is some dimension-one parameter and $M$ is a cutoff-size parameter. During its evolution, $\phi$ takes a range of values in such a way that the effective masses of the Higgs bosons are scanned, starting from the values of order $M$, and finishing close to the zero point. The stop of the scanning is provided by the potential barrier arising shortly after one of the Higgses acquires a negative mass and hence a non-zero VEV, by the mechanism described below.  
The scalar potential realizing the above idea is the following
\bea \label{eq:potential}
V &=& (|h_1|^2+|h_2|^2) (M^2 - g \phi)  + \lambda (|h_1|^4+|h_2|^4)  \nn \\
&& - g M^2 \phi + \dots  \nn \\
&& + m^2 (|h_1|^2 - |h_2|^2) \cos {\phi \over f} + {\kappa m^4 \over (4 \pi)^2} \cos^2 {\phi \over f},
\eea
where the dots stand for subleading terms, which will not affect our discussion. 
The first line of Eq.~(\ref{eq:potential}) contains a $Z_2$ symmetric potential of $h$-bosons and their interactions with $\phi$, as described in the beginning of this section. The second line drives the $\phi$ field evolution, from $\phi < M^2 /g$ (by assumption) to $\phi \sim M^2 /g$, where the Higgs mass squared crosses zero.  The potential barriers are generated by the terms in the last line of Eq.~(\ref{eq:potential}). 

We assume that the potential~(\ref{eq:potential}) features several approximate symmetries. First of all, the scanning field has a shift symmetry $\phi \to \phi +c$, which is completely broken by the terms proportional to the parameter $g$, and also partially broken down to $\phi \to \phi +2 \pi n$ by terms proportional to $m$.   
Hence we will be able to set these two dimension-one parameters much smaller than the cutoff scales of our description $M$ and $f$ without conflict with naturalness. However a simultaneous presence of both types of breakings puts doubts on a possibility to UV complete such type of models within the framework of local quantum field theory, see~\cite{Gupta:2015uea,Batell:2015fma} for a discussion.
The second approximate symmetry is the aforementioned $Z_2$. By its virtue the cutoff-enhanced corrections to the term $m^2 (|h_1|^2 - |h_2|^2) \cos {\phi \over f}$, appearing when one closes Higgs loops, get cancelled. Instead, the only additional oscillating terms are proportional to $m^4$, as is shown explicitly in the last line of Eq.~(\ref{eq:potential}), where $\kappa$ is a dimensionless coefficient $\log$-sensitive to the cutoff. 
 
Let us now consider the details of the cosmological evolution of our model. The field $\phi$ will roll down its potential, releasing the excess of its energy by Hubble friction. The latter requires that the relaxation mechanism proceeds and ends before the end of inflation.  To start with, we will discuss the evolution of $\phi$ and $h$-fields assuming that the fundamental parameters of the potential~(\ref{eq:potential}) remain constant, and later on we will comment on the possible effects of time-dependence related to the change of the inflaton field value.   

While the $\phi$ field rolls down its potential, the vacuum expectation values of the $h$-fields, when non-zero, will be
\be \label{eq:v12}
v^2_{1,2}={g \phi - M^2 \mp m^2   \cos {\phi \over f} \over 2 \lambda} \equiv {m^2 (p \mp  \cos {\phi \over f}) \over 2 \lambda},
\ee
where we have introduced the parameter $p = {(g \phi - M^2) / m^2}$, which, as we will see later, describes the large-scale evolution of $\phi$, in contrast to $\phi$ explicitly appearing in cosine, describing shorter-scale oscillations. 
In the following we will also make use of the fact that in the SM $\lambda \ll 1$.

For the early times, when $p<-1$, both Higgs VEV's are equal to zero and in order to ensure that the relaxion passes this region without stop we need to impose the condition $V^{\prime}_{\phi}<0$, hence from Eq.~(\ref{eq:potential})
\be \label{eq:reg_1}
  g M^2 > {\kappa m^4 \over (4 \pi)^2 f}.  
\ee 
At the late times, when $p>1$, both VEV's become non-zero, hence according to Eq.~(\ref{eq:v12}) their difference is fixed, $v_1^2-v_2^2=-m^2/\lambda \cos \phi/f$. Therefore the relaxion potential acquires EWSB-dependent oscillatory contribution with a constant amplitude $\sim m^4/\lambda$. If these oscillations have a sufficient magnitude to produce the barriers, the relaxion will be stopped somewhat before the $p=1$ point, because after $p=1$ all the barriers have the same height. In order to trap $\phi$ we need the condition $V^{\prime}_{\phi}>0$ to be satisfied for $p>1$, hence 
\be \label{eq:reg_3}
 g M^2 < {m^4 \over 2 \lambda f}.  
\ee
From Eq.~(\ref{eq:v12}), taking $p\sim 1$ one can estimate the size of the Higgs VEV's to be at most $v_{1,2}^2\sim m^2/\lambda$, which allows to make the observed small EW scale technically natural. It is worth noticing that in our model trapping the relaxion in a minimum with $v,m_h\ll M$ requires that {\it a priori} independent symmetry breaking parameters $g$ and $m$, besides being small, are related to each other so that Eqs~(\ref{eq:reg_1}), (\ref{eq:reg_3}) are satisfied. This type of coincidence, pointed out in~\cite{Espinosa:2015eda},  can not be explained within the considered EFT, but pose no problem from the technical naturalness point of view.     

Using Eq.~(\ref{eq:reg_3}) and requiring the new physics scale to be above the electroweak scale, $M^2\gg M_W^2 \sim m^2$, we get $g f \ll M_W^2,m^2$. Therefore the step between two minima of the potential $\delta \phi \sim \pi f$ corresponds to the change of the Higgs mass parameter by $g \delta \phi \sim g f \ll M_W^2$. In this regard, the parameter $p$ can be considered as a parametrization for the large-scale growth of $\phi$, while the explicit dependence on $\phi$ in the expressions above gives small-scale oscillations. 

Now let us consider more closely the intermediate regime $|p|<1$, where one can expect that the amplitude of the oscillations gradually grows until it gets saturated at $p=1$. As follows from $g f \ll m^2$, moving from $p=-1$ to $p=1$ takes many periods of $\cos \phi/f$ oscillations. 
Depending on the relative size of $p$ and $\cos \phi/f$, one or both $h$-fields will periodically get non-vanishing VEV's:    


\begin{itemize}

\item
$p< -|\cos {\phi \over f}|$: no VEV's  and hence no barriers, provided~(\ref{eq:reg_1}) holds.

\item
$- |\cos {\phi \over f}| < p <  |\cos {\phi \over f}|$: only one Higgs VEV is different from zero. Substituting expressions for VEV's from Eq.~(\ref{eq:v12})
into  Eq.~(\ref{eq:potential}) we find that in order to have the extrema of the $\phi$ potential one has to satisfy the condition
\be \label{eq:condit_2}
{m^4\over 2 \lambda f }\left(\pm p + \cos {\phi \over f}\right)\sin {\phi \over f} = g M^2 \;\;\;
\text{for} \;\;\; \cos {\phi \over f} \gtrless 0 \,.
\ee

\item
$p >  |\cos {\phi \over f}|$: both Higgs fields will have non-vanishing VEV's and the $V^{\prime}_{\phi} = 0$ condition reads
\be \label{eq:condit_3}
{m^4\over 2 \lambda f }  \sin {2 \phi \over f} = g M^2\,.
\ee
At the same time requiring these extrema to be maxima we get $\cos 2\phi > 0$ and hence $p > 1/\sqrt 2$. It is trivial to show that starting from the point $p = 1/\sqrt 2$ the maximal slope of the potential does not increase anymore.

\begin{figure}
\centering
\includegraphics[width=.45\textwidth]{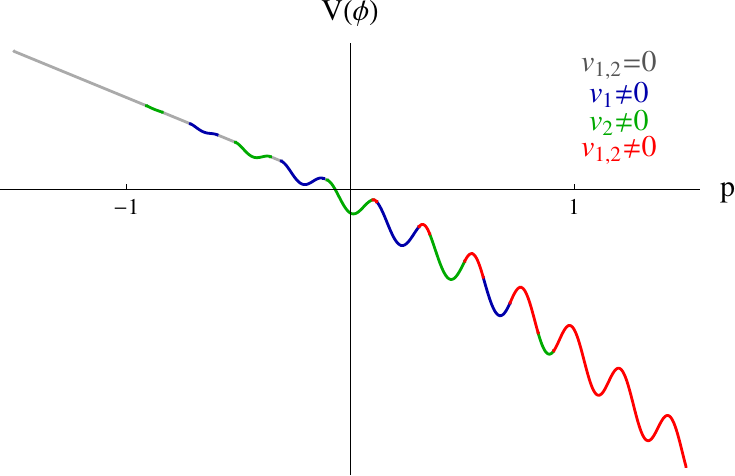}
\hfill
\includegraphics[width=.45\textwidth]{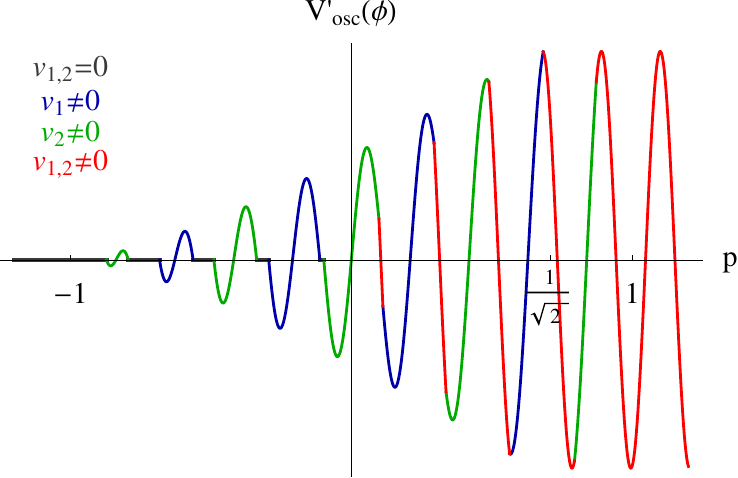}
\caption{Left panel: schematic representation of the potential of the scanning field $\phi$, reflecting three stages of its evolution. First ($p<-1$) both $h$-fields have zero VEV's. Then ($|p|<1$) one or both of them get a non-zero VEV periodically, this is where the potential barriers have to stop $\phi$. In the third region ($p>1$) both VEV's are non-zero, and their difference is fixed, producing the $\phi$ oscillations of the constant amplitude. Colors reflect the values of the Higgs VEV's as explained on the plot. Here we neglected subdominant EWSB-independent oscillations.
Right panel: the slope generated by the oscillatory terms of the potential, i.e. excluding the contribution from $g M^2 \phi$ in Eq.~(\ref{eq:potential}).}
\label{fig:phipotential}
\end{figure}

\renewcommand{\labelitemi}{$\blacktriangleright$}

\end{itemize}

The above considerations mean the following. Starting from $p=-1$ up to $p=1/\sqrt 2$ the amplitude of the oscillatory term grows gradually, from ${\kappa m^4 / 16 \pi^2 f}$ to ${m^4 / 2 \lambda f}$. In all the local minima on this way only one of the $h$-VEV's is non-vanishing. The minima with both VEV's being non-zero can appear only starting from  the point $p=1/\sqrt 2$, when the amplitude reaches its absolute maximum. This implies that the condition $V^\prime=0$, allowing the $\phi$ field to stop the scanning, will typically be satisfied for the first time in the region with only one non-zero VEV.  This will happen for      
\be \label{eq:summary_stop}
 {\kappa m^4 \over 16 \pi^2 f} < g M^2 < {m^4 \over 2 \lambda f}\,,
\ee
trapping the relaxion in a vacuum with $v^2_{1} \simeq {m^2 \over 2 \lambda}$ and $v_2 = 0$. More precisely, the value of the remaining non-zero VEV can be found from Eq.~(\ref{eq:v12}), substituting there the minimal value of $p$ satisfying Eq.~(\ref{eq:condit_2}) together with a corresponding value of $\phi$. 
Though the relaxion is not expected to strictly follow the classical evolution described above, the quantum spreading would not allow it to move far from the classical stopping point, provided sufficiently small Hubble scale during inflation as will be discussed later.  
We schematically present the main features of the relaxion potential and its slope on Fig.~\ref{fig:phipotential}.

So far we were assuming no relevant variation of the fundamental parameters of the potential~(\ref{eq:potential}) with a time, which led us to the conclusion that after relaxation both SM's will have similar Higgs masses, but one of the Higgs VEV's is necessarily zero. As will be discussed in the following, this scenario can be phenomenologically viable, but it is also worth analysing the ways to make both VEV's non-vanishing. 
One can for instance consider a possibility that both Higgses couple to one of the inflation sector fields in a $Z_2$-symmetric way, such that the evolution of the inflation sector, happening after the relaxion has settled in its minimum, can induce additional shifts to the Higgs mass squared parameters and make both negative. Though the Higgs-inflaton couplings are generically present~\cite{Gross:2015bea}, it needs further investigations whether and how they can be used for the mentioned above purpose without spoiling the inflation itself. 
Nevertheless in the remaining part of this paper we will be considering both possible situations, with one and two non-zero Higgs VEV's.

The final comment is that one can add to the potential~(\ref{eq:potential}) a term $\lambda_{12} |h_1|^2|h_2|^2$ linking directly the two sectors. This term will not bring any qualitative changes to the relaxation mechanism, but will have some phenomenological consequences which we will comment on later. One can estimate the minimal size of $\lambda_{12}$ from the quantum corrections it gets, $\lambda_{12} \gtrsim m^4 / f^4$.

\section{Consistency of Scanning Mechanism}
\label{sec:consist}

Here we review the conditions which are necessary for the mirror relaxation mechanism to work within the inflationary epoch. All but the last one were formulated in~\cite{Graham:2015cka} and~\cite{Espinosa:2015eda}, therefore we refrain from their detailed discussion.

{\it a) Unwanted quantum corrections should be kept small.} For instance the periodic part of the potential after closing the loops with $\phi$ field would generate $\phi$-independent corrections to $h_1$ and $h_2$ masses, breaking the $Z_2$ symmetry
\be
m_{h_1}^2-m_{h_2}^2  \sim  {m^2 \over 16 \pi^2} {M^2 \over f^2},
\ee 
which has to be smaller than the typical size of Higgs mass at the final stage of relaxation, i.e. $m^2$, hence we derive a constraint $M\lesssim f$ \footnote{We thank Giuliano Panico for pointing out this constraint.}.

{\it b) by assumption the inflation is not disturbed by the $\phi$-$h_{1,2}$ sector}, hence its overall vacuum energy density variation, $\sim M^4$, should be less than the energy density driving the inflation $V_{i}\sim M_{\text{Pl}}^2 H_i^2$, so 
\be \label{eq:cond_2}
H_i > {M^2 \over M_{\text {Pl}}}\,,
\ee
where $M_{\text{Pl}}$ is a reduced Planck mass and $H_i$ is a Hubble scale during inflation.
In addition, the $\phi$ field time variation should always be subdominant with respect to $V_i$, 
\bea
&{\dot \phi^2 \over V_i} \simeq \left( { V^\prime_\phi \over H_i^2 M_{\text{Pl}} } \right)^2 = \left( {g M^2 \over H_i^2 M_{\text{Pl}}}\right)^2 \ll 1& \,, \\
&{\ddot \phi \over 3 H_i \dot \phi} \simeq {V^{\prime \prime}_\phi \over H_i^2} = {g^2 \over H_i^2} \ll1 \,, &
\eea 
This ensures the slow-roll regime for $\phi$, in which $\dot \phi \sim V^\prime_\phi / H_i$.

{\it c) consistency of the description in terms of classical rolling} requires quantum jumps of the scanning field, $\sim H_i$, to be subdominant with respect to the classical slow-roll displacement, hence
\be \label{eq:cond_3}
H_i < (g M^2)^{1\over 3}\,.
\ee

{\it d) scanning over a sufficient range of the $\phi$ field values}, $\Delta \phi \sim M^2/g$ puts a constraint on the number of e-folds of inflation
\be \label{eq:cond_4}
N_e > {H_i^2 \over g^2}\,.
\ee

{\it e) relaxion does not step over the barriers} hence $H_i < m$. 

{\it f)}
{\it stability against reheating} is important if the reheating temperature is above the electroweak scale. 
As we will see in Sec.~\ref{sec:thermhist}, only one type of particles (SM$_1$ or SM$_2$) 
is allowed to be produced upon reheating. Then the temperature corrections can lead to a restoration of the electroweak symmetry in the reheated sector. The other sector representatives 
will not be produced and not enter thermal equilibrium with the first sector. Hence the potential barriers for the $\phi$ field, produced by the first Higgs, will disappear, while the ones associated with the second Higgs 
will still be present. Therefore the $\phi$ field will not be able to roll further than one period ahead and most likely remain in the same region ($v_{1/2}\ne0$ or $v_{1,2}\ne0$) it was before. In case if it is the region which features just one non-zero VEV, after the temperature comes down, the system will appear in the vacuum where the reheated SM will have a zero VEV. Hence the reheating with a temperature above the electroweak scale necessarily requires the post-relaxational adjustment of the Higgs VEV's as was described in Section~\ref{sec:main} or some other sort of additional corrections to the Higgs VEV's to make both non-zero.

{\it g)}
{\it  }
The variation of the Higgs mass in the potential~(\ref{eq:potential}) induced by the inflaton evolution should not outrun the $\phi$-induced shift. The inflaton-induced shift generated from the time of trapping to the end of inflation, if positive, has to be small. If negative, it should be at most of the size of the electroweak scale and can possibly be used to set the value of the latter.

Combining the constraints~(\ref{eq:cond_2}),(\ref{eq:cond_3}) with the trapping condition~(\ref{eq:summary_stop}) we obtain
\be
M^2 f^{1/3} < (m v)^{2/3} M_{\text{Pl}}\,,
\ee
therefore for $m=m_h$ and $f\sim M$ we obtain $M <  10^9\text{GeV}$, and for $f\sim 10^{-2} M_{\text{Pl}}$ one gets $M < 10^8\text{GeV}$.

\section{Thermal History}
\label{sec:thermhist}

Cosmological data puts severe bounds on the theories containing mirror copies of the Standard Model. Primarily, these are the constraints on the energy density of new light degrees of freedom present during the Big Bang Nucleosynthesis (BBN)~\cite{Iocco:2008va} and Cosmic Microwave Background (CMB)~\cite{Brust:2013xpv} formation epochs. Since in our construction the mirror Higgs VEV is either zero, or of the SM Higgs VEV size, the total number of light degrees of freedom (below $\sim$MeV scale) in both sectors will be at least doubled compared to the SM. Hence the presence of both types of matter with equal abundances in the observed part of the Universe is ruled out by BBN and CMB constraints.
In the following we will sketch two possible mechanisms allowing to satisfy aforementioned bounds by dumping the density of mirror states. 

\begin{itemize}

\item  
The first type of mechanisms~\cite{Kolb:1985bf,Hodges:1993yb,Berezinsky:1999az} assumes that each SM copy is coupled to its own inflation sector, and the two reheatings happen at different time. Hence after one of inflatons reheats its SM$_1$, the other continues driving the inflation which dilutes the already produced SM$_1$ states. After that, the second inflaton reheats the remaining SM$_2$. 
The crucial feature of our model is a weakly broken $Z_2$ symmetry, which may be violated if the two inflatons take very different values.    
In order to preserve this symmetry during relaxation, one could for instance consider a combined inflation (e.g.~\cite{Dvali:2003vv}), in which at the first stage both inflatons remain in a $Z_2$ symmetric vacuum, until the end of the electroweak relaxation. Afterwards the inflatons consecutively tunnel to another region of the potential where they evolve, continuing triggering inflation, and eventually reheat the corresponding sectors, with a dilution of the sector which was reheated first.  
This mechanism can work both with two or with just one non-vanishing Higgs VEV's, but in the latter case the zero-VEV SM will form the dominant type of matter in a half of the space patches of the Universe.

\item
The second idea is to use the difference in microscopical features of the two sectors~\cite{Berezhiani:1995am}. In our particular scenario this difference is the most pronounced if one of the Higgses does not acquire VEV.  If the characteristic scales defining the SM state production upon reheating are not higher than the electroweak scale, production of the ordinary matter can be enhanced by $v\ne0$.
As an example let us consider an inflaton which is lighter than the Higgs bosons and couples to them by means of interactions $\sim \sigma (|h_1|^2+|h_2|^2)$. The field $\sigma$ may also be some other light field the inflaton predominantly decays to. In this case the reheating of the SM and its copy will be mediated by off-shell Higgs bosons. Reheating of the SM states will mostly proceed via $\sigma$ decays to two bottom quarks, while one of the dominant mirror states production channels would be $\sigma$ decaying to four almost massless mirror tops. 
 Hence the mirror states production will be dumped by the phase space of four-body decay. The ratio of $\sigma$ decay widths to mirror and ordinary states is
\be
{\widetilde \Gamma \over \Gamma} \sim {1 \over (2\pi)^6} {m_\sigma^6 \over m_h^4 m_b^2} \sim 10^{-3} \left(m_\sigma \over 100 \text{GeV} \right)^6 \,,
\ee
where here and in the following the tilded symbols correspond to the mirror sector. 
This suppression translates into the ratio of two sectors temperatures, ${\widetilde T_R / T_R} \sim ( {\widetilde \Gamma /  \Gamma} )^{1/2}$\cite{Linde:2005ht}. If the two sectors are decoupled, the temperature ratio  will not change significantly by the BBN and CMB times. Hence the contribution of the mirror degrees of freedom to the effective number of light species in the periods of interest will be $\delta N_\nu \sim ~\tilde g_{\star} ({\widetilde T_R / T_R})^4 \sim 10^{-4} (m_\sigma / 100 \text{GeV})^{12}$~\cite{Brust:2013xpv}. 
While the current experimental data allows for $\delta N_\nu\sim10^{-1}$~\cite{Ade:2015xua} deviations from the SM prediction.

\end{itemize}

A common feature of both these dumping mechanisms is a requirement that the reheating of the ordinary sector should not be followed by a sizeable production of the mirror states. For this the expansion rate of the Universe should be greater than the mirror states production rate.
In case if the maximal post-inflationary temperatures are above the electroweak scale, the dominant production of the mirror states is  $h_1 h_1 \to h_2 h_2$ mediated by the $\phi$ field or the quartic interaction $\lambda_{12} |h_1|^2 |h_2|^2$. The rate of this process can be estimated as 
\be
\Gamma_{2h_1 \to 2h_2}  = n \langle \sigma v \rangle \sim \lambda_{12}^2 T\,,
\ee
while for the $\phi$-mediated production one substitutes  $\lambda_{12} \to m_h^4 / f^2 T^2$, where $T$ is temperature.
Requiring this reaction rate to be smaller than the Hubble parameter $H\sim g_\star^{1/2} T^2 / M_{\text{Pl}}$, where $g_\star \sim 100$, we get~\cite{Lew:1993af}
\be
 f > 10^6 \text{GeV}  \;\;\;\;\text{and}  \;\;\;\;
\lambda_{12} < 10^{-8} \,,
\ee
where for the temperature we took the minimal value allowing the reaction to proceed, $T\sim m_h$, which puts the strongest constraint.
These bounds can be relaxed if after inflation the temperature was never higher than the electroweak scale~\cite{Ignatiev:2000yy}.  
If both sectors have non-vanishing Higgs VEV's of a similar size, one can expect that the dominant $\widetilde{\text{SM}}$ production (for $T\sim 10$~GeV) reaction is $bb \to \tilde b \tilde b$, mediated by the off-shell Higgses. The process rate can be estimated as 
\be
\Gamma_{2b \to 2\tilde b} \sim  {T^5 m_b^4 \lambda_{12}^2 \over m_h^8}
\ee
 for the quartic interaction and for the $\phi$-mediated process we substitute $\lambda_{12} \to m_h^4 / f^2 T^2$.  
This would lead to constraints
\be
f > 10^5 \left({T\over \text{GeV}}\right)^{-1/4} \text{GeV}  \;\;\;\;\text{and}  \;\;\;\;
\lambda_{12} < 10^{-2} \left({T\over \text{GeV}}\right)^{-3/2} \,.
\ee
If instead the mirror VEV is zero, one of the dominant production modes is $2 b \to 4 \tilde t$. An additional phase space suppression further decreases the reaction rate
\be
\Gamma_{2b \to 4 \tilde t} \sim  {T^{11} m_b^2 \lambda_t^2 \lambda_{12}^2 \over 4 (2 \pi)^6 m_h^{12}}\,,
\ee
which leads to the bounds
\be
f > 500 \left({T\over 10 \, \text{GeV}}\right)^{5/4} \text{GeV}  \;\;\;\;\text{and}  \;\;\;\;
\lambda_{12} < 10 \left({T\over 10 \, \text{GeV}}\right)^{-9/2} \,.
\ee
Obtained in this way limits show that in principle the interactions between the two sectors can be sufficient to allow for the observable effects in the current and future collider experiments.  Though there is no reason for the new physics effects not to be suppressed to a negligible size, in contrast to for example twin-Higgs models~\cite{Chacko:2005pe,Craig:2015pha,Barbieri:2015lqa,Low:2015nqa}.

\section{Discussion}
\label{sec:disc}

The proposal of~\cite{Graham:2015cka} for a dynamical solution of the Higgs mass problem opened a new avenue for the development of the beyond Standard Model physics scenarios. In this note we described one such a scenario, allowing to address the gauge hierarchy problem and move the cutoff to the scales up to $10^9$~GeV.   Our model does have new physics at the electroweak scale, represented by the mirror SM, but it can be negligibly weakly coupled to the ordinary Standard Model and hence unobserved in current or even far future collider experiments.
The described construction has the following features. The masses of the Higgs bosons in the two sectors are typically of the same order.  However in the minimal case one of the Higgs VEV's equals zero, making the mirror SM similar to the Higgsless SM as described for instance in~\cite{Quigg:2009xr}. Provided by an appropriate inflation plus reheating model one may also expect to have both Higgs VEV's to be non-zero.

Given the presence of the mirror SM, one has to require the post-inflationary Universe to be populated only with the ordinary matter, for which we sketched some possibilities.  In case of one inflationary sector coupled to both copies of the SM, the viable physics can be obtained when one of the Higgs VEV's is vanishing (this is ensured automatically in the simplest case of our model) and the inflaton mass is not above the EW scale.   
In case of two separate inflationary sectors, the $v_{1/2}\ne0$ realization can work only if the reheating temperature $T_{\text R}$ is lower than the EW scale, while the $v_{1,2}\ne0$ case allows for $T_{\text R}>M_{W}$.
  
Regarding the subsequent cosmological evolution, the scenario we considered can be split into two cases. In the first case the postinflationary reheating temperature is above the electroweak scale, hence the communication between the two sectors is restricted to be almost vanishing. For the low $T_{\text R}$, instead, the mirror SM can in principle have sizeable couplings to the ordinary SM, opening a possibility for a direct detection of the former. 

The two sectors can interact directly, or via interactions with the scanning field.
The scanning field $\phi$ can have a very light mass, $\sim m_h^2 / f$, which is $10$~KeV for $f = 10^9$~GeV. Being very weakly coupled to SM$_{1,2}$ states it can be cosmologically stable, though a careful estimate of it viability as a DM candidate needs a separate discussion. One may also be able to obtain dark matter candidates from the mirror SM if it is not completely diluted (see~\cite{Barbieri:2005ri,Khlopov1,Khlopov2,Khlopov3,Khlopov4,Khlopov5,Khlopov6} for works in the context of more general mirror-symmetric models). 

The mechanism of the mirror relaxation, the possibility of post-relaxational adjustment of the Higgs sector parameters, dumping of the mirror matter production during the evolution of the Universe  strongly depend on the details of particular inflation+reheating models, which we  ignored in this note. Hence a dedicated analysis of the inflationary sector would be important.

\subsection*{Acknowledgments}

We thank Riccardo Barbieri, Christopher Murphy, Giuliano Panico and David Pirtskhalava for many useful discussions. This work was supported by MIUR under the contract 2010 YJ2NYW-010 and in part by the MIUR-FIRB grant RBFR12H1MW.


\end{document}